\documentstyle[12pt]{article}
\newcommand{\blankline}{\vskip .3cm}
\newcommand{\f}{\begin{equation}}
\newcommand{\ff}{\end{equation}}
\newcommand{\be}{\begin{equation}}
\newcommand{\ee}{\end{equation}}
\newcommand{\bea}{\begin{eqnarray}}
\newcommand{\eea}{\end{eqnarray}}
\begin{document}
\centerline{\LARGE ``Junk'' DNA as a genetic decoy} \blankline
\blankline \rm \centerline{Jo\~ao Magueijo}\blankline
\centerline{\it Theoretical Physics Group, The Blackett
Laboratory, Imperial College of Science, Technology and Medicine }
\centerline{\it South Kensington, London SW7 2BZ, UK}



\blankline \blankline \blankline \blankline \centerline{}
\blankline \blankline \blankline \blankline

It is well known \cite{mirk,th,tc} that in most species a large
proportion of the genome does not appear to code genes or
proteins, or even regulate coding DNA. This leads to the
descriptive expression ``junk DNA''. Naturally, it is possible
that the hidden information in this DNA has simply escaped us so
far; however that such sequences have high redundancy (they may,
for example, be long repetitions of the same base) gives weight to
their characterization as junk. This raises the question: if junk
DNA is indeed useless, what might be its evolutionary advantage?

An obvious answer is that there is {\it no} evolutionary advantage
to junk DNA; however its presence doesn't cause any harm either.
In particular the extra metabolic activity it implies may be
negligible, because the percentage of metabolic activity
associated with DNA replication is very small. Since only
features harmful to survival or reproduction are selected out, a
neutral accessory such as junk DNA could still survive.

This is certainly a factor \cite{crick}, and in some insects it
has been shown that the genome size is related to the ability to
eliminate DNA. However, for it to be the complete answer to this
question one would need to prove that eliminating junk DNA by
chance (i.e. by random mutation) carries a suitably low
probability in all species. Also it is not always true that the
energy cost associated with DNA replication is
negligible~\cite{cd}.

A much more interesting possibility is that junk DNA may be a
positive evolutionary feature, even though it serves no useful
direct purpose. In this short note we suggest that its role is to
protect the gene. Most significantly its presence may baffle the
actions of mutation agents, such as retro-viruses. In the presence
of large amounts of junk,  a random genetic agent hits junk most
of the time; the meaningful pieces of the genome meanwhile
``hide'' in the foliage. The higher the percentage of junk  the
higher the protection it affords to directly useful DNA.

The validity of this argument is far from general. It must
certainly be part of the story regarding retroviral insertions,
which can indeed be found within junk DNA (along with the
low-information sequences mentioned above). According to our
hypothesis such insertions are missed hits. However the argument
is not so clearcut (and in some cases {\it has} to break
down) with regards to environmental mutagens. Then junk DNA 
only stands a chance of protecting the gene if junk sequences wool around
the low density important sequences. Even then the argument does
not apply, for instance, to ionizing radiation. If a harmful
photon or cosmic ray crosses $n$ junk DNA sections before crossing
a ``good'' one, and if $p\ll 1$ is the probability of interaction
with any one of them, then the probability of a harmful mutation
is $p(1-p)^n\approx p$, i.e. the junk did nothing to decrease the
cross-section. (The reason why this argument does not apply to
retoviruses is that $p$ is not small once the virus penetrates
within striking distance of the DNA). So we are not claiming that
junk DNA protects the gene against a general mutation; only that
this is a factor in some cases.

With this proviso, the hypothesis that junk DNA protects the gene
may be tested by considering that the {\it percentage} of junk DNA
varies widely from specie to specie, with {\it Drosophila}
possessing one of the lowest \cite{ch} and the salamander one of
the highest. This allows us to predict a number of possible
correlations. Let us consider first the effect of retro-viruses.
If junk DNA acts as a decoy {\it only} against unwanted
retro-viral insertions, then there should be a close link between
the number of different retro-viruses attacking a given specie and
its percentage of junk DNA (the fact that some junk DNA are viral
insertions will induce a feedback complication to this link). The
higher the threat, the higher the protection.

However, other factors interfere with this correlation. Although
viral insertions are generally bad news, mutations are also
useful, supplying natural selection with variety. But mutation is
only beneficial as long as it does not over-run the ability of a
species to stabilize its positive features. The ideal mutation
rate depends directly on how prolifically a given species
reproduces and how short its cycles are. The percentage of junk
DNA should reflect this optimized mutation rate.

If a large part of the offspring goes to waste anyway, there is no
harm is having a lot of mutation in every generation. This is the
case for most insects. If on the contrary a species reproduces
very little, and it is crucial to preserve a large proportion of
the offspring, then mutation has to be severely suppressed, as the
species can only rarely afford a random mutation. This is the case
for large mammals. Thus there should be an anti-correlation
between the reproductive ability of a species and the percentage
of junk DNA it contains (this is consistent with known estimates).
More directly, there should be an anti-correlation between the
natural mutation rates of a species and its percentage of junk
DNA.

There certainly are many other factors, together with the number
of retro-viruses and the natural ideal mutation rates, determining
the level of protection that should be given to useful DNA in a
given species. But the point we wish to make in this note is that,
whatever the answer, it is likely that junk DNA is an expression
of this delicate balance. That junk DNA percentages are at best
rough estimates makes a direct comparison futile at this stage.
But as our understanding of genomes improves,  it should be
possible to test this idea.

\blankline

{\bf  ACKNOWLEDGMENTS} I would like to thank R. Badawi, K.
Baskerville, J. Brosius, C. Cantor, and L. Pogosian, for
discussions. I also thank two anonymous referees for helpful
comments (and for pointing out the obvious: that I am not an
expert in the field). I am grateful for D. Bosanac for organizing
the interdisciplinary conference ``Space, time, and life''
(Brijuni, 2002), where the idea in this paper was born.

\end{document}